\documentclass{ws-ijmpcs}
\usepackage{amsmath}
\usepackage{amsfonts}
\usepackage{amssymb}

\begin{document}

\markboth{M. Hohmann}{Non-metric fluid dynamics and cosmology on Finsler spacetimes}

\catchline{}{}{}{}{}
%
%%%%%%%%%%%%%%%%%%%%%%%%%%%%%%%%%%%%%%%%%%%%%%%%%%%%%%%%%%%%%%%%%%%%

\title{Non-metric fluid dynamics and cosmology on Finsler spacetimes}

\author{Manuel Hohmann}

\address{Laboratory of Theoretical Physics, Institute of Physics, University of Tartu,\\
Ravila 14c, 50411 Tartu, Estonia\\
manuel.hohmann@ut.ee}

\maketitle

\begin{abstract}
We generalize the kinetic theory of fluids, in which the description of fluids is based on the geodesic motion of particles, to spacetimes modeled by Finsler geometry. Our results show that Finsler spacetimes are a suitable background for fluid dynamics and that the equation of motion for a collisionless fluid is given by the Liouville equation, as it is also the case for a metric background geometry. We finally apply this model to collisionless dust and a general fluid with cosmological symmetry and derive the corresponding equations of motion. It turns out that the equation of motion for a dust fluid is a simple generalization of the well-known Euler equations.
\end{abstract}

\ccode{PACS numbers: 05.20.Dd, 02.40.Hw, 04.50.Kd, 98.80.Jk}

\section{Motivation}\label{sec:motivation}
Modern cosmology has confronted the physics community with a number of unexplained observations, such as the accelerating expansion of the universe and the isotropy of the cosmological microwave background. The standard model of cosmology aims to explain these observations by the presence of dark energy in the $\Lambda$CDM model and an inflationary phase in the early universe. However, the mechanisms behind these explanations, i.e., the constituents of dark energy and the driving force of the inflation, are not yet understood. This opens the possibility to create various models both from particle physics and gravity theory. The model we discuss here belongs to the gravitational category. The basic idea of this model is to replace the metric geometry of spacetime from general relativity by a Finsler length measure. In this article we derive a theory of fluid dynamics based on this Finsler geometric background.

The work we present in this article has two main ingredients. The first ingredient, which serves as the background geometry, is the Finsler spacetime framework.\cite{Pfeifer:2011tk,Pfeifer:2011xi,Pfeifer:2013gha,Pfeifer:2014yua} This framework introduces a notion of Finsler geometry which provides a unified description of a Lorentzian causality, observers and gravity, and which can be used as a background for field theories such as electrodynamics. The second ingredient is the kinetic theory of fluids.\cite{Ehlers:1971,Sarbach:2013fya,Sarbach:2013uba} This theory is based on the idea that fluids can be modeled as being constituted by particles, whose worldlines are geodesics on the background spacetime. In the continuum limit this theory yields equations of motion for a density function on a subspace of the tangent bundle of spacetime. In this article we generalize this kinetic theory of fluids to Finsler spacetimes. In this model the density function becomes a function on observer space,\cite{Gielen:2012fz,Hohmann:2013fca,Hohmann:2014gpa,Hohmann:2015pva} which is the space of physically allowed four-velocities. We apply this formalism to two physically motivated examples.

The outline of this article is as follows. In section~\ref{sec:finsler} we provide a brief introduction to the geometry of Finsler spacetimes. From this starting point we construct the space of physical observer velocities in section~\ref{sec:observer} and discuss its geometric structure. We then use this structure to construct a model of fluid dynamics based on kinetic theory in section~\ref{sec:fluids}. In section~\ref{sec:special} we apply this model to two particular examples: a collisionless dust fluid and a fluid with cosmological symmetry. We end with a conclusion in section~\ref{sec:conclusion}.

\section{Finsler spacetime geometry}\label{sec:finsler}
We start our discussion with a brief review of the background geometry. In this section we introduce the basic geometric objects on a Finsler spacetime, which will later be used for the construction of fluid dynamics. Our starting point is the following definition:\cite{Pfeifer:2011tk,Pfeifer:2011xi,Pfeifer:2013gha}

A \emph{Finsler spacetime} \((M,L,F)\) is a four-dimensional, connected, Hausdorff, paracompact, smooth manifold \(M\) equipped with continuous real functions \(L, F\) on the tangent bundle \(TM\) which has the following properties:
\begin{enumerate}
\item\label{finsler:lsmooth}
\(L\) is smooth on the tangent bundle without the zero section \(TM \setminus \{0\}\).
\item\label{finsler:lhomogeneous}
\(L\) is positively homogeneous of real degree \(n \geq 2\) with respect to the fiber coordinates of~\(TM\),
\begin{equation}
L(x,\lambda y) = \lambda^nL(x,y) \quad \forall \lambda > 0\,,
\end{equation}
and defines the Finsler function \(F\) via \(F(x,y) = |L(x,y)|^{\frac{1}{n}}\).
\item\label{finsler:lreversible}
\(L\) is reversible: \(|L(x,-y)| = |L(x,y)|\).
\item\label{finsler:lhessian}
The Hessian
\begin{equation}
g^L_{ab}(x,y) = \frac{1}{2}\bar{\partial}_a\bar{\partial}_bL(x,y)
\end{equation}
of \(L\) with respect to the fiber coordinates is non-degenerate on \(TM \setminus X\), where \(X \subset TM\) has measure zero and does not contain the null set \(\{(x,y) \in TM | L(x,y) = 0\}\).
\item\label{finsler:timelike}
The unit timelike condition holds, i.e., for all \(x \in M\) the set
\begin{multline}
\Omega_x = \Bigg\{y \in T_xM \Bigg| |L(x,y)| = 1,\\
g^L_{ab}(x,y) \text{ has signature } (\epsilon,-\epsilon,-\epsilon,-\epsilon), \epsilon = \frac{L(x,y)}{|L(x,y)|}\Bigg\}
\end{multline}
contains a non-empty closed connected component \(S_x \subseteq \Omega_x \subset T_xM\).
\end{enumerate}

Here we have used coordinates \((x^a)\) on \(M\) and the \emph{induced coordinates} \((x^a,y^a)\) on \(TM\) such that \((x,y) = y^a\partial_a|_x \in T_xM\). The Finsler function \(F\) introduced above defines a length functional
\begin{equation}\label{eqn:finslerlength}
s[\gamma] = \int d\tau\,F(\gamma(\tau),\dot{\gamma}(\tau))\,,
\end{equation}
which measures the length of a curve \(\tau \mapsto \gamma(\tau)\) on \(M\). An important class of curves called \emph{Finsler geodesics} are those for which the length functional becomes extremal. They satisfy the geodesic equation
\begin{equation}\label{eqn:geodesic}
\ddot{\gamma}^a + N^a{}_b(\gamma,\dot{\gamma})\dot{\gamma}^b = 0\,.
\end{equation}
Here we have introduced the \emph{Cartan non-linear connection}
\begin{equation}\label{eqn:nonlinconn}
N^a{}_b = \frac{1}{4}\bar{\partial}_b\left[g^{F\,ac}(y^d\partial_d\bar{\partial}_cF^2 - \partial_cF^2)\right]\,,
\end{equation}
where \(\partial_a = \partial/\partial x^a\), \(\bar{\partial}_a = \partial/\partial y^a\) and \(g^{F\,ab}\) is the inverse of the \emph{Finsler metric}
\begin{equation}
g^F_{ab}(x,y) = \frac{1}{2}\bar{\partial}_a\bar{\partial}_bF^2(x,y)\,.
\end{equation}
The Cartan non-linear connection induces a unique split of the tangent bundle \(TTM\) into horizontal and vertical parts, \(TTM = HTM \oplus VTM\). The horizontal tangent bundle \(HTM\) is spanned by the vector fields
\begin{equation}
\{\delta_a = \partial_a - N^b{}_a\bar{\partial}_b\}\,,
\end{equation}
while the vertical tangent bundle \(VTM\) is spanned by \(\{\bar{\partial}_a\}\). The corresponding basis \(\{\delta_a,\bar{\partial}_a\}\) that respects this split is called the \emph{Berwald basis}. Its dual basis of \(T^*TM\) is given by
\begin{equation}
\{dx^a, \delta y^a = dy^a + N^a{}_bdx^b\}\,.
\end{equation}
Using the Berwald basis we can find another description for Finsler geodesics. For this purpose we consider the canonical lift \(\Gamma: \mathbb{R} \to TM\) of a geodesic \(\gamma\) to the tangent bundle \(TM\). Writing \(\Gamma\) in coordinates \((\Gamma^a,\bar{\Gamma}^a) = (\gamma^a,\dot{\gamma}^a)\) we find the geodesic equation
\begin{equation}
\dot{\Gamma}^a = \bar{\Gamma}^a\,, \quad \dot{\bar{\Gamma}}^a = -N^a{}_b(\Gamma^a,\bar{\Gamma}^a)\bar{\Gamma}^b\,.
\end{equation}
Since this is a first order differential equation, the canonical lifts of Finsler geodesics take the form of integral curves of a vector field \(\mathbf{S} = y^a\delta_a\) on \(TM\), which we call the \emph{geodesic spray}. Finally, the Berwald basis and the Finsler metric allow the construction of a metric
\begin{equation}\label{eqn:sasakimetric}
G = -g^F_{ab}\,dx^a \otimes dx^b - \frac{g^F_{ab}}{F^2}\,\delta y^a \otimes \delta y^b\,.
\end{equation}
on the tangent bundle, which is called the \emph{Sasaki metric}.

\section{Observer space geometry}\label{sec:observer}
After discussing the most important Finsler geometric structures on the tangent bundle \(TM\) in the previous section we now restrict ourselves to a particular subspace of \(TM\). Recall from the definition of a Finsler spacetime that for each \(x \in M\) there exists a shell \(S_x \subset T_xM\) of future unit timelike vectors, which corresponds to the four-velocities of test masses and physical observers. Their union
\begin{equation}\label{eqn:observerspace}
O = \bigcup_{x \in M}S_x
\end{equation}
is called \emph{observer space}. This is the space on which we will define fluid dynamics in the remainder of this work.

The observer space is a seven-dimensional submanifold of the tangent bundle \(TM\) and thus inherits a number of geometric structures from the Finsler geometry. In particular, the geodesic spray \(\mathbf{S}\) introduced in the previous section is tangent to the observer space, and thus restricts to a vector field \(\mathbf{r}\) on \(O\), called the \emph{Reeb vector field}. This relation between the geodesic spray and the observer space is in fact necessary for the interpretation of \(O\) as the space of physical four-velocities: it means that a test mass possessing a physical initial four-velocity and following a geodesic, and thus an integral curve of \(\mathbf{S}\), will have a physical four-velocity at all times. The canonical lifts of all physical geodesics, corresponding to freely falling test masses, are thus given by integral curves of the Reeb vector field \(\mathbf{r}\) on \(O\).

The second important structure on the observer space is the restriction \(\tilde{G}\) of the Sasaki metric \(G\) to \(O\). This also equips \(O\) with a volume form \(\Sigma = \mathrm{Vol}_{\tilde{G}}\). This volume form has a number of properties which are relevant for the construction of a theory of fluid dynamics. The most relevant property for the remainder of this article is the fact that it is preserved by the flow of the Reeb vector field, \(\mathcal{L}_{\mathbf{r}}\Sigma = 0\).

From the Reeb vector field \(\mathbf{r}\) and the volume form \(\Sigma\) we finally define the \emph{particle measure} \(\omega = \iota_{\mathbf{r}}\Sigma\). It is the unique six-form (up to a constant factor) which has the following properties necessary for the construction of fluid dynamics from the kinetic theory. Most importantly, its restriction to a hypersurface \(\sigma \subset O\) which is not tangent to the Reeb vector field \(\mathbf{r}\) is a non-vanishing volume form on \(\sigma\). Further, it is closed, \(d\omega = 0\), and preserved by the flow of the Reeb vector field, \(\mathcal{L}_{\mathbf{r}}\omega = 0\). The relevance of these properties will become clear in the following section, when we use it as an ingredient to develop a theory of fluid dynamics.

\section{Kinetic theory and fluid dynamics}\label{sec:fluids}
We now turn our attention to the kinetic theory of fluids\cite{Ehlers:1971,Sarbach:2013fya,Sarbach:2013uba} and apply it to fluids on the Finsler spacetime background geometry detailed in the preceding section. For this purpose we first briefly review how fluids can be modeled by the geodesic motion of particles. From the geodesic motion on Finsler spacetimes we then derive the equations of motion for a Finsler fluid.

The basic idea of the kinetic theory of fluids is the assumption that fluids are constituted by particles. In the simplest possible case of a single component fluid, which we will discuss here, all particles have identical properties, such as mass and electric charge, and follow piecewise geodesic curves. This geodesic motion corresponds to the motion of freely falling test masses without any other interaction. The interaction between particles is modeled by collisions, which mark the endpoints of the geodesic pieces of the particle trajectories and correspond to instantaneous transfers of momentum between the particles. The physical background of this model of interactions is the assumption that the interaction distances are small compared to the distances between the particles and that the interaction times are short compared to the time between interactions, which is the case for a sufficiently low density.

In order to construct a continuum theory of fluids from the particle model one introduces the one-particle distribution function \(\phi: O \to \mathbb{R}^+\) such that for each oriented hypersurface \(\sigma \subset O\) the integral
\begin{equation}
N[\sigma] = \int_{\sigma}\phi\omega
\end{equation}
is the number \(N[\sigma]\) of particle trajectories whose canonical lifts to \(O\) pass through \(\sigma\). Here particle trajectories \(\gamma\) are counted positively (negatively) if the tangent vector of their canonical lift and a positively oriented basis of the tangent space to \(\sigma\) at the intersection point between \(\gamma\) and \(\sigma\) form a positively (negatively) oriented basis of the tangent space of \(O\) at that point. Further, \(\omega\) denotes the particle measure introduced in the previous section.

We now take a closer look at the canonical lifts of the particle trajectories to the observer space. It follows from our assumption of instantaneous momentum transfer that these lifts are discontinuous at collisions, i.e., they have endpoints corresponding to the particles' velocities before and after the collision. To incorporate collisions into the continuum theory we define the collision transfer density \(\dot{\phi}: O \to \mathbb{R}\) such that for each hypervolume \(V \subset O\) the integral
\begin{equation}
\dot{N}[V] = \int_V\dot{\phi}\Sigma
\end{equation}
is the number of initial points minus the number of final points of canonical lifts of particle trajectories in \(V\). Here \(\Sigma\) denotes the volume form of the Sasaki metric \(\tilde{G}\) on \(O\).

Note that we have defined the counting prescription for curves passing a hypersurface such that any curve which has an initial point, but no final point in the hypervolume \(V\) passes its boundary \(\partial V\) in the positive direction, and in the negative direction in the opposite case. It thus follows that \(\dot{N}[V] = N[\partial V]\). From this we further find that
\begin{equation}
\int_V\dot{\phi}\Sigma = \dot{N}[V] = N[\partial V] = \int_{\partial V}\phi\omega = \int_Vd(\phi\omega) = \int_V(\mathcal{L}_{\mathbf{r}}\phi)\Sigma\,,
\end{equation}
where we used Stokes' theorem and the properties of the particle measure \(\omega\). Since this holds for any hypervolume \(V\) it follows that
\begin{equation}
\dot{\phi} = \mathcal{L}_{\mathbf{r}}\phi\,.
\end{equation}
The right hand side of this equation describes the evolution of \(\phi\). We thus obtain the equation of motion of the fluid by specifying a functional \(\dot{\phi}[\phi]\) describing collisions between particles. For the simplest possible case of a collisionless fluid we have \(\dot{\phi} = 0\), and the fluid equation of motion reduces to the Liouville equation \(\mathcal{L}_{\mathbf{r}}\phi = 0\).

This concludes our discussion of the kinetic theory of fluids in general. In the following section we will discuss particular examples of fluids and derive their equations of motion.

\section{Special cases}\label{sec:special}
After discussing general fluids on Finsler spacetimes in the previous section, we now consider a few special cases. The first case we display in section~\ref{subsec:dust} is dust, which is conventionally described as a collisionless perfect fluid with vanishing pressure. Starting from the Liouville equation we derive the generalization of the Euler equations to dust on Finsler spacetimes. The second case shown in section~\ref{subsec:cosmology} is the most general cosmological fluid. We start from a general Finsler spacetime with cosmological symmetry and derive the equations of motion for the most general fluid obeying the same symmetry.

\subsection{Dust fluid}\label{subsec:dust}
The first example we discuss is a collisionless dust fluid characterized by a matter density \(\rho(x)\) and four-velocity distribution \(u^a(x)\), which is a future timelike vector field normalized by the Finsler function, and thus a function \(u: M \to O\). The aim of this section is to derive the equations of motion for \(\rho\) and \(u\) from the kinetic theory introduced in the preceding section. For this purpose we will first construct the one-particle distribution function \(\phi\) and then impose the Liouville equation \(\mathcal{L}_{\mathbf{r}}\phi = 0\) for a collisionless fluid.

We start by introducing coordinates \((\hat{x}^a,\theta^{\alpha}) = (x^a,y^{\alpha}/y^0)\) on \(O\) and writing the three-velocity as \(v^{\alpha} = u^{\alpha}/u^0\). Together with the proper Dirac delta distribution on the unit timelike shell \(S_{\hat{x}}\) we find the one-particle distribution function
\begin{equation}\label{eqn:dustphi}
\phi(\hat{x},\theta) = \frac{1}{m}\rho(\hat{x})\frac{\delta(\theta - v(\hat{x}))}{\sqrt{h^F(\hat{x},\theta)}}\,,
\end{equation}
where \(h^F\) denotes the determinant of the restriction
\begin{equation}
h^F_{\alpha\beta} = \frac{\partial y^a}{\partial\theta^{\alpha}}\frac{\partial y^b}{\partial\theta^{\beta}}g^F_{ab}
\end{equation}
of the Sasaki metric to \(S_{\hat{x}}\). A direct calculation of the Lie derivative \(\mathcal{L}_{\mathbf{r}}\phi\) is obstructed by the presence of the delta distribution. We therefore introduce an arbitrary smooth test function \(f: O \to \mathbb{R}\). Using the properties of the particle measure \(\omega\) and integrating by parts we then find
\begin{equation}
0 = \int_{O}f\left(\mathcal{L}_{\mathbf{r}}\phi\right)\Sigma = \int_{O}d(\phi\omega)f = -\int_{O}df \wedge \omega\phi = -\int_{O}d(f\omega)\phi =
-\int_{O}\phi\left(\mathcal{L}_{\mathbf{r}}f\right)\Sigma\,.
\end{equation}
Writing the Finsler function in the form
\begin{equation}
F = y^0\tilde{F}(x,y/y^0) = y^0\tilde{F}(\hat{x},\theta)
\end{equation}
we obtain the Reeb vector field
\begin{equation}
\mathbf{r} = y^a\left[\hat{\partial}_a + \tilde{F}\left(N^0{}_a\theta^{\alpha} - N^{\alpha}{}_a\right)\tilde{\partial}_{\alpha}\right],
\end{equation}
using the partial derivatives \(\hat{\partial}_a\) and \(\tilde{\partial}_{\alpha}\) with respect to \(\hat{x}^a\) and \(\theta^{\alpha}\), and the volume form
\begin{equation}
\Sigma = \sqrt{g^Fh^F}d^4\hat{x}d^3\theta\,.
\end{equation}
Finally inserting the explicit form for \(\phi\) we arrive at
\begin{equation}
0 = \int_Od^4\hat{x}d^3\theta\sqrt{g^F}\delta(\theta - v)\rho y^a\left[\hat{\partial}_a + \tilde{F}\left(N^0{}_a\theta^{\alpha} - N^{\alpha}{}_a\right)\tilde{\partial}_{\alpha}\right]f\,.
\end{equation}
Integrating over \(d^3\theta\) cancels the Dirac distribution and yields
\begin{equation}
0 = \left.\int_Md^4\hat{x}\sqrt{g^F}\rho u^a\left[\hat{\partial}_af + \tilde{F}\left(N^0{}_av^{\alpha} - N^{\alpha}{}_a\right)\tilde{\partial}_{\alpha}f\right]\right|_{y^a = u^a(\hat{x})}\,,
\end{equation}
where it is indicated that all objects on observer space are taken at the point \((\hat{x},v(\hat{x}))\). Observe that the partial derivative \(\hat{\partial}_af(\hat{x},v(\hat{x}))\) is fixed by \(f(\hat{x},v(\hat{x}))\) and thus not an independent quantity, while \(\tilde{\partial}_{\alpha}f(\hat{x},v(\hat{x}))\) depends on the choice of \(f\) in a neighborhood of \(v(\hat{x})\), which can be chosen independently. We therefore need to eliminate \(\hat{\partial}_af\) using integration by parts. Since \(\hat{\partial}_a\) is only a partial derivative, but \(f(\hat{x},v(\hat{x}))\) contains also an implicit dependence on \(\hat{x}\) via \(v(\hat{x})\), we first need to rewrite the partial derivative into a total derivative using
\begin{equation}
\hat{\partial}_af(\hat{x},v(\hat{x})) = \frac{d}{d\hat{x}^a}f(\hat{x},v(\hat{x})) - \hat{\partial}_av^{\alpha}\tilde{\partial}_{\alpha}f(\hat{x},v(\hat{x}))\,.
\end{equation}
Integration by parts and using the normalization \(u^0\tilde{F}(\hat{x},v(\hat{x})) = F(\hat{x},u(\hat{x})) = 1\) then finally yields
\begin{multline}
0 = \int_Md^4\hat{x}\sqrt{g^F}\bigg\{\left[\frac{u^a}{u^0}\left(N^0{}_a\frac{u^{\alpha}}{u^0} - N^{\alpha}{}_a\right) - u^a\hat{\partial}_a\left(\frac{u^{\alpha}}{u^0}\right)\right]\rho\tilde{\partial}_{\alpha}f\\
- \left[\hat{\partial}_a(\rho u^a) + \frac{1}{2}\rho u^ag^{F\,bc}\left(\hat{\partial}_ag^F_{bc} + \hat{\partial}_au^d\bar{\partial}_dg^F_{bc}\right)\right]f\bigg\}\bigg|_{y^a = u^a(\hat{x})}\,.
\end{multline}
Since now all spacetime derivatives act on objects which depend only on spacetime coordinates, we can rename the coordinates \(\hat{x}^a\) back to \(x^a\) and read off the equations of motion
\begin{equation}\label{eqn:dusteomraw1}
u^a\left(N^0{}_a\frac{u^{\alpha}}{u^0} - N^{\alpha}{}_a\right) - u^a\partial_au^{\alpha} + \frac{u^au^{\alpha}}{u^0}\partial_au^0 = 0
\end{equation}
and
\begin{equation}\label{eqn:dusteomraw2}
\partial_a(\rho u^a) + \frac{1}{2}\rho u^ag^{F\,bc}\left(\partial_ag^F_{bc} + \partial_au^d\bar{\partial}_dg^F_{bc}\right) = 0\,.
\end{equation}
These equations can further be simplified. Using the properties of the Cartan non-linear connection we can rewrite the first equation~\eqref{eqn:dusteomraw1} in the form
\begin{equation}\label{eqn:dusteom1}
0 = u^a(\partial_au^b + N^b{}_a) = \nabla u^b\,,
\end{equation}
where we have introduced the dynamical covariant derivative \(\nabla\). Similarly, the second equation~\eqref{eqn:dusteomraw2} can be written as
\begin{equation}\label{eqn:dusteom2}
0 = \partial_a(\rho u^a) + \frac{1}{2}\rho u^ag^{F\,bc}\delta_ag^F_{bc} = \nabla_{\delta_a}(\rho u^a)\,,
\end{equation}
where we have introduced the covariant derivative \(\nabla_{\delta_a}\) of the Cartan linear connection. In the case of a metric Finsler function \(F^2(x,y) = |g_{ab}(x)y^ay^b|\) these equations reduce to the well-known Euler equations
\begin{equation}
u^b\nabla_bu^a = 0 \quad \text{and} \quad \nabla_a(\rho u^a) = 0
\end{equation}
for a pressureless fluid.

\subsection{Cosmological fluid}\label{subsec:cosmology}
As the final example we will derive the Liouville equation for a fluid with cosmological symmetry. For this purpose we introduce coordinates \((t,r,\vartheta,\varphi)\) on the spacetime manifold \(M\) and the corresponding induced coordinates \((t,r,\vartheta,\varphi,y^t,y^r,y^{\vartheta},y^{\varphi})\) on \(TM\). In these coordinates the cosmological symmetry generators take the form
\begin{gather}
\xi_1 = \sqrt{1 - kr^2}\left(\sin\vartheta\cos\varphi\partial_r + \frac{\cos\vartheta\cos\varphi}{r}\partial_{\vartheta} - \frac{\sin\varphi}{r\sin\vartheta}\partial_{\varphi}\right)\,,\nonumber\\
\xi_2 = \sqrt{1 - kr^2}\left(\sin\vartheta\sin\varphi\partial_r + \frac{\cos\vartheta\sin\varphi}{r}\partial_{\vartheta} + \frac{\cos\varphi}{r\sin\vartheta}\partial_{\varphi}\right)\,,\nonumber\\
\xi_3 = \sqrt{1 - kr^2}\left(\cos\vartheta\partial_r - \frac{\sin\vartheta}{r}\partial_{\vartheta}\right)\,, \quad
\xi_6 = \partial_{\varphi}\,,\nonumber\\
\xi_4 = \sin\varphi\partial_{\vartheta} + \frac{\cos\varphi}{\tan\vartheta}\partial_{\varphi}\,, \quad
\xi_5 = -\cos\varphi\partial_{\vartheta} + \frac{\sin\varphi}{\tan\vartheta}\partial_{\varphi}\,,\label{eqn:cosmovect}
\end{gather}
where \(k \in \{-1,0,1\}\) determines the spatial curvature of the corresponding spacetime. For cosmologically symmetric fluid dynamics we require that both the background geometry and the one-particle distribution function of the fluid obey this symmetry.

We start by deriving the most general cosmologically symmetric background geometry. A Finsler spacetime is symmetric under the action of a vector field \(\xi\) if and only if the geometry function \(L\), and thus also the Finsler function \(F\), is invariant under the complete lift of \(\xi\) to the tangent bundle,\cite{Pfeifer:2011xi,Pfeifer:2013gha,Hohmann:2015pva}
\begin{equation}
\left(\xi^a\partial_a + y^b\partial_b\xi^a\bar{\partial}_a\right)F = 0\,.
\end{equation}
In the present case of a cosmological symmetry it is most convenient to introduce new coordinates on \(TM\) via the definition
\begin{gather}
t = \hat{t}\,, \quad r = \hat{r}\,, \quad \vartheta = \hat{\vartheta}\,, \quad \varphi = \hat{\varphi}\,, \quad y^t = \hat{y}\,,\nonumber\\
y^r = \hat{w}\cos\hat{u}\sqrt{1 - k\hat{r}^2}\,, \quad y^{\vartheta} = \frac{\hat{w}}{\hat{r}}\sin\hat{u}\cos\hat{v}\,, \quad y^{\varphi} = \frac{\hat{w}}{\hat{r}\sin\hat{\vartheta}}\sin\hat{u}\sin\hat{v}\,.
\end{gather}
After calculating the complete lifts of the vector fields~\eqref{eqn:cosmovect} in these coordinates it turns out that the most general Finsler function with cosmological symmetry is given by \(F = F(\hat{t},\hat{y},\hat{w})\). Further, from the fact that \(F\) is 1-homogeneous in the coordinates \(\hat{y}\) and \(\hat{w}\) follows that it must be of the form
\begin{equation}
F(\hat{t},\hat{y},\hat{w}) = \hat{y}\tilde{F}\left(\hat{t},\hat{w}/\hat{y}\right)\,.
\end{equation}
For the discussion of geodesic motion it is convenient to introduce yet another set of coordinates on \(TM\) as
\begin{equation}
\tilde{t} = \hat{t}\,, \quad \tilde{r} = \hat{r}\,, \quad \tilde{\vartheta} = \hat{\vartheta}\,, \quad \tilde{\varphi} = \hat{\varphi}\,, \quad \tilde{u} = \hat{u}\,, \quad \tilde{v} = \hat{v}\,, \quad \tilde{y} = \hat{y}\tilde{F}\left(\hat{t},\frac{\hat{w}}{\hat{y}}\right)\,, \quad \tilde{w} = \frac{\hat{w}}{\hat{y}}\,.
\end{equation}
In these coordinates the observer space \(O\) is given as a connected component of the submanifold \(\tilde{y} = 1\), so that one can use the remaining seven coordinates to parametrize \(O\). The Reeb vector field then takes the form
\begin{multline}
\mathbf{r} = \frac{1}{\tilde{F}}\Bigg(\tilde{\partial}_t + \tilde{w}\cos\tilde{u}\sqrt{1 - k\tilde{r}^2}\tilde{\partial}_r + \frac{\tilde{w}\sin\tilde{u}\cos\tilde{v}}{\tilde{r}}\tilde{\partial}_{\vartheta} + \frac{\tilde{w}\sin\tilde{u}\sin\tilde{v}}{\tilde{r}\sin\tilde{\vartheta}}\tilde{\partial}_{\varphi}\\
- \frac{\tilde{w}\sin\tilde{u}\sqrt{1 - k\tilde{r}^2}}{\tilde{r}}\tilde{\partial}_u - \frac{\tilde{w}\sin\tilde{u}\sin\tilde{v}}{\tilde{r}\tan\tilde{\vartheta}}\tilde{\partial}_v - \frac{\tilde{F}_{tw}}{\tilde{F}_{ww}}\tilde{\partial}_w\Bigg)\,,
\end{multline}
where the subscripts \(t\) and \(w\) indicate derivatives with respect to \(\tilde{t}\) and \(\tilde{w}\), respectively.

We finally come to the discussion of fluids on the Finsler spacetime background derived above. Since the background geometry obeys the cosmological symmetry defined by the vector fields~\eqref{eqn:cosmovect}, their canonical lifts are tangent to the observer space \(O\). A fluid obeys the same symmetry if and only if its one-particle distribution function \(\phi\) is invariant under the restriction of these canonical lifts to \(O\). In the present case the most general one-particle distribution function satisfying this condition takes the form \(\phi = \phi(\tilde{t},\tilde{w})\). Its Lie derivative with respect to the Reeb vector field is thus given by
\begin{equation}
\mathcal{L}_{\mathbf{r}}\phi = \frac{1}{\tilde{F}}\left(\phi_t - \frac{\tilde{F}_{tw}}{\tilde{F}_{ww}}\phi_w\right)\,.
\end{equation}
For the simplest possible case of a collisionless fluid the equation of motion hence takes the form
\begin{equation}
\tilde{F}_{ww}\phi_t = \tilde{F}_{tw}\phi_w\,.
\end{equation}
This is the Liouville equation for a fluid with cosmological symmetry.

\section{Conclusion}\label{sec:conclusion}
We have derived a model for fluid dynamics on Finsler geometric backgrounds based on the kinetic theory of fluids and applied our model to two important special cases. Our results show that Finsler spacetimes provide a suitable background geometry for fluid dynamics and that the obtained fluid equations of motion reduce to their well-known limits if the background geometry is metric.

Of course any model of fluid dynamics must be complemented by a suitable model of gravitational dynamics in order to derive consistent solutions. For the Finsler spacetimes we used here an extension of general relativity exists.\cite{Pfeifer:2011xi,Pfeifer:2013gha} The source of the gravitational field in this model is a scalar function on the tangent bundle of spacetime. Deriving this energy-momentum scalar for a kinetic fluid and the corresponding gravitational field equations will be a topic of future research.

\section*{Acknowledgments}
The author is happy to thank Christian Pfeifer for extensive discussions and fruitful collaboration. He gratefully acknowledges the full financial support of the Estonian Research Council through the Postdoctoral Research Grant ERMOS115 and the Startup Research Grant PUT790.

\end{document}